\documentclass[aps,prc,showpacs,showkeys,twoside,twocolumn,byrevtex,floatfix]{revtex4-2}

\usepackage[dvips]{graphicx} 
\usepackage{color}           
\usepackage{amsmath,amssymb,amsfonts}
\usepackage{bm}
\usepackage{mathtools}

\newcommand{\nc}{\newcommand}           
\nc{\vc}[1]     {\mbox{\boldmath $#1$}} 
\nc{\mapleft}[1]{                       
 \smash{\mathop{                      %
  \hbox to 0.90cm{\rightarrowfill} }\limits_{#1}}}

\nc{\beq}     {\begin{eqnarray}}
\nc{\eeq}    {\end{eqnarray}}
\nc{\bra}       {\langle}               
\nc{\ket}       {\rangle}               
\nc{\bras}[1]   {\langle#1|}            
\nc{\kets}[1]   {|#1\rangle}            
\nc{\del}       {\partial}              

\newcommand{\lw}[1]{\smash{\lower1.75ex\hbox{#1}}}

\nc{\red}[1]    {\textcolor{black}{#1}}  

\nc{\mydraft}	{\setlength{\topmargin}{-1.5cm}}
\mydraft
\begin{document}

\title{
  Shell and cluster structures in $^{20}$Ne in the variation of multiple bases of the antisymmetrized molecular dynamics 
}

\author{Takayuki Myo}
\email{takayuki.myo@oit.ac.jp}
\affiliation{General Education, Faculty of Engineering, Osaka Institute of Technology, Osaka, Osaka 535-8585, Japan}
\affiliation{Research Center for Nuclear Physics (RCNP), Osaka University, Ibaraki, Osaka 567-0047, Japan}

\author{Mengjiao Lyu}
\affiliation{College of Science, Nanjing University of Aeronautics and Astronautics, Nanjing 210016, China}

\author{Qing Zhao}
\affiliation{School of Science, Huzhou University, Huzhou 313000, Zhejiang, China}

\author{Masahiro Isaka}
\affiliation{Faculty of Intercultural Communication, Hosei University, Chiyoda-ku, Tokyo 102-8160, Japan}

\author{Niu Wan}
\affiliation{School of Physics and Optoelectronics, South China University of Technology, Guangzhou 510641, China}

\author{Hiroki Takemoto}
\affiliation{Faculty of Pharmacy, Osaka Medical and Pharmaceutical University, Takatsuki, Osaka 569-1094, Japan} 
\affiliation{Research Center for Nuclear Physics (RCNP), Osaka University, Ibaraki, Osaka 567-0047, Japan}

\author{Hisashi Horiuchi}
\affiliation{Research Center for Nuclear Physics (RCNP), Osaka University, Ibaraki, Osaka 567-0047, Japan}

\author{Akinobu Dot\'e}
\affiliation{KEK Theory Center, Institute of Particle and Nuclear Studies (IPNS), High Energy Accelerator Research Organization (KEK), Tsukuba, Ibaraki, 305-0801, Japan}
\affiliation{J-PARC Branch, KEK Theory Center, IPNS, KEK, Tokai, Ibaraki, 319-1106, Japan}

\date{\today}

\begin{abstract}%
  We investigate the structures of $^{20}$Ne in the variation of the multiple bases of the antisymmetrized molecular dynamics (AMD).
  In this method, the multiple AMD bases are superposed and optimized simultaneously in the total-energy variation.
  This scheme is beneficial for describing the various configurations in $^{20}$Ne.
  In the results, we confirm the shell and cluster structures in the $K^\pi=0^+_{1-4}$ bands,
  such as the deformed states in the $K^\pi=0^+_{1,4}$ bands with the $\alpha$ cluster development,
  and the spherical shell-like states in the $K^\pi=0^+_2$ band,
  the latter of which is difficult to describe in the previous AMD calculations imposing the quadrupole deformation. 
  We evaluate the monopole and quadrupole transitions in these states.
  The negative parity states of $^{20}$Ne with $K^\pi=0^-$ and $2^-$ are discussed in relation to the shell and cluster structures.
  As a result, six kinds of the $K^\pi$ bands in $^{20}$Ne are described comprehensively in the microscopic framework of nuclei.
\end{abstract}

\maketitle

\section{Introduction}

Nuclear clustering is one of the major properties of nuclei as well as the shell-model states and the mean-field states
\cite{ikeda68,horiuchi12,freer18}.
In the cluster states, some nucleons form clusters, such as an $\alpha$ particle, and they are spatially developed in nuclei.
A typical case is a $^8$Be nucleus, which decays into two $\alpha$ particles.
The cluster states are often observed near the threshold energy of the $\alpha$ particle emission
because of the weak interaction between clusters. This property is called the ``threshold rule'' \cite{ikeda68}.

The $^{20}$Ne nucleus is one of the $4n$ self-conjugate nuclei with $N=Z$, and there have been many studies on the clustering aspect
of this nucleus \cite{fujiwara80}.
The thresholds of the cluster emission are located in the low-excitation energy region of $^{20}$Ne;
$^{16}$O+$\alpha$ is 4.73 MeV and $^{12}$C+$2\alpha$ is 11.89 MeV, which are lower than
the threshold energies of the one-proton (one-neutron) emission of 12.84 MeV (16.86 MeV), respectively.
It is interesting to investigate the clustering effect in the low-lying states of $^{20}$Ne together with the shell-model and mean-field aspects.
Recently, the candidates of the $5\alpha$ condensate state have been examined experimentally \cite{adachi21} and theoretically \cite{zhou23}.

Experimentally, the $K^\pi=0^+_{1-4}$ bands are known as well as the $K^\pi=0^-$ and $2^-$ bands \cite{tilley98,tunl,nndc}.
The $K^\pi=0^\pm$ bands are considered to be the parity doublets \cite{fujiwara80}.

In the shell-model scheme, the $K^\pi=0^+_{1,2}$ band states are well described in the $(sd)^4$ configuration space
upon the $^{16}$O core \cite{tomoda78};
however, the $K^\pi=0^+_{3,4}$ band states are difficult to describe because of the possible $\alpha$ cluster emergence in these states.

\red{In Refs. \cite{kimura04,isaka11}, the authors apply the nuclear model of the antisymmetrized molecular dynamics (AMD) to $^{20}$Ne.}
The AMD wave function is a Slater determinant consisting of the Gaussian wave packet for a nucleon, and
the spatial positions of wave packets are responsible for describing the cluster states as well as the shell-like states. 
\red{The authors also introduce the deformation of the Gaussian wave packet to treat the deformed mean-field effect.}
They generate the AMD basis states under the constraints on the quadrupole deformation
and successfully describe the $K^\pi=0^+_{1,4}$, $0^-$, and $2^-$ band states. 
However, the $K^\pi=0^+_{2,3}$ band states are not obtained in this approach.
In the mean-field model \cite{marevic18}, the $K^\pi=0^+_1$ and $0^-$ band states are also discussed. 

Experimentally, the $K^\pi=0^+_3$ band states are known to be populated by the $^8$Be transfer reaction on $^{12}$C
\cite{middleton71,nagatani71,greenwood75}.
\red{Considering this situation, in Ref. \cite{taniguchi04},
the authors adopt the constraints of the $^{12}$C+2$\alpha$ configurations for $^{20}$Ne in the AMD model.}
They successfully describe the $K^\pi=0^+_3$ band states, which contain the $^{12}$C+2$\alpha$ configurations to some extent;
however, the $K^\pi=0^+_2$ band states are not obtained.

\red{
On the theory side, the recent progress of the {\it ab initio} calculations for $^{20}$Ne
has been presented in the symmetry-adapted no-core shell model \cite{tobin14} and the lattice effective field theory \cite{lahde14}.
}

As mentioned above, the $K^\pi=0^+_2$ band is considered to be the excited states in the shell-model configurations \cite{tomoda78},
but this state is difficult to obtain in the usual energy variation utilizing the quadrupole deformation
in the microscopic methods such as AMD and the mean-field model.
\red{
  In particular, in the previous AMD calculations \cite{kimura04,isaka11,taniguchi04},
  the authors superpose the basis states with various quadrupole deformations, but the $0^+_2$ band is not obtained.
}
It is demanded to introduce a different approach, in which one can generate efficient basis states
suitable to describe the $K^\pi=0^+_2$ band of $^{20}$Ne together with other $K^\pi$ bands.

In relation to this demand, we have recently developed a new scheme,
in which the variation of the multiple configurations is performed using the AMD basis states \cite{myo23b}.
The multiple AMD configurations are superposed and each configuration is optimized simultaneously with respect to the energy variation of the total system.
We also extend the method to generate the excited-state configurations imposing the orthogonal condition on the ground-state configurations.
In the previous works \cite{myo23b,myo25,tian24,cheng24,tian25}, we have applied this method to light nuclei and hypernuclei, and
have discussed the emergence of the cluster states as well as the shell-like ones, in relation to the threshold rule for the emission of clusters.

In AMD, the energy variation is called the cooling method (the imaginary-time evolution)
and the present extension is the cooling for the multiple AMD bases, namely the multiple cooling, or simply abbreviated as the ``multicool method.''
The advantage of the multicool method is that one does not suppose the physical constraints, such as radius and deformation,
to optimize the multiple configurations.
This method is considered to be promising in the description of the $K^\pi=0^+_2$ band states with the shell-like states in $^{20}$Ne.

In this study, we apply the multicool method to the AMD wave function for $^{20}$Ne
and discuss the various structures of $^{20}$Ne comprehensively.
In particular, we focus on the $K^\pi=0^+_{1-4}$, $0^-$ and $2^-$ bands.

In Sec.~\ref{sec:method}, we explain the formulation of the variation of the multiple Slater determinants in the AMD wave function.
In Sec.~\ref{sec:result}, we discuss the results of $^{20}$Ne. 
In Sec.~\ref{sec:summary}, we summarize this work.

\section{Theoretical methods}\label{sec:method}

\subsection{Antisymmetrized molecular dynamics (AMD)}\label{sec:AMD}

We explain the AMD model for nuclei \cite{kimura04,isaka11,taniguchi04}.
The AMD wave function $\Phi_{\rm AMD}$ is a Slater determinant of the $A$-nucleon system given as
\begin{eqnarray}
  \begin{split}
\Phi_{\rm AMD}
&= \frac{1}{\sqrt{A!}} {\rm det} \left\{ \prod_{i=1}^A \phi_i(\bm{r}_i) \right\} ,
\\
\phi_i(\bm{r})&=\left(\frac{2\nu}{\pi}\right)^{3/4} e^{-\nu(\bm{r}-\bm{Z}_i)^2} \chi_{\sigma,i} \chi_{\tau,i} ,
\\
\chi_{\sigma,i} &= \alpha^\uparrow_i \kets{\uparrow} + \alpha^\downarrow_i \kets{\downarrow}.
  \end{split}
\label{eq:AMD}
\end{eqnarray}
The single-nucleon wave function $\phi_i(\bm{r})$ has a Gaussian wave packet with a common range parameter $\nu$, where $\hbar \omega=2\nu \hbar^2/m$ with a nucleon mass $m$,
and the individual centroid parameters $\bm{Z}_i$.
The condition of $\sum_{i=1}^A \bm{Z}_i={\bf 0}$ is imposed to set the mean position of the center of mass of $\Phi_{\rm AMD}$ to be the origin.
The spin part $\chi_{\sigma}$ is expressed in terms of the up and down components with the weights of $\alpha^{\uparrow/\downarrow}_i$,
and the isospin part $\chi_{\tau}$ is a proton or a neutron.

The energy minimization with the Hamiltonian $H$ is performed in the following cooling equations
for total energy $E^\pm_{\rm AMD}$ with parity projection $P^\pm$.
Using the arbitrary negative number $\mu$, the parameters $X_i=\{\bm{Z}_i,\alpha^{\uparrow/\downarrow}_i\}$ are determined:
\begin{equation}
  \begin{split}
    \Phi^\pm_{\rm AMD}&= P^\pm \Phi_{\rm AMD},
    \\
    E^\pm_{\rm AMD} &= \dfrac{ \bra \Phi^\pm_{\rm AMD}|H| \Phi^\pm_{\rm AMD} \ket }{\bra \Phi^\pm_{\rm AMD}| \Phi^\pm_{\rm AMD} \ket},
    \\
    \dfrac{{\rm d} X_i}{{\rm d} t}&= \frac{\mu}{\hbar} \dfrac{\partial E^\pm_{\rm AMD}}{ \partial X_i^*},\quad \mbox{and c.c}.
  \end{split}
  \label{eq:cooling}
\end{equation}
Using $\Phi_{\rm AMD}$, the angular momentum projection with the operator $P^J_{MK}$ is performed for the spin state $(J,M,K)$:
\begin{eqnarray}
  \begin{split}
  \Psi^{J^\pm}_{MK,{\rm AMD}}
  &= P^J_{MK}P^{\pm} \Phi_{\rm AMD}.
  \end{split}
  \label{eq:projection}
\end{eqnarray}

\red{
  The AMD wave function is extendable to the multiple configurations with a number $N_{J^\pm}$,
  where the basis states are often generated with the constraints on the nuclear deformation \cite{horiuchi12,kimura04}.
  The total wave function $\Psi_{\rm t}^{J^\pm}$ is a superposition of the projected AMD basis states in Eq.~(\ref{eq:projection}),
denoted as $\Psi_n^{J^\pm}$ simply:
\begin{eqnarray}
  \begin{split}
   \Psi_{\rm t}^{J^\pm}
&= \sum_{n=1}^{N_{J^\pm}} C_n^{J^\pm}\,  \Psi_n^{J^\pm} \,, \quad
   E_{\rm t}^{J^\pm} = \dfrac{ \bra \Psi_{\rm t}^{J^\pm} |H| \Psi_{\rm t}^{J^\pm} \ket }{\bra \Psi_{\rm t}^{J^\pm}| \Psi_{\rm t}^{J^\pm} \ket}, 
  \end{split}
   \label{eq:linear}
\end{eqnarray}
where $n$ is the index of the projected AMD basis states with $J^\pm$ including the $K$ mixing.
From the variational principle of the total energy, $\delta E_{\rm t}^{J^\pm}=0$, the generalized eigenvalue problem, namely the Hill-Wheeler equation,
is solved to obtain $\Psi_{\rm t}^{J^\pm}$ and $E_{\rm t}^{J^\pm}$:
\begin{eqnarray}
  \begin{split}
   \sum_{n=1}^{N_{J^\pm}} \Bigl\{ \langle\Psi_{m}^{J^\pm} | H |\Psi_{n}^{J^\pm}\rangle - E_{\rm t}^{J^\pm} \langle\Psi_{m}^{J^\pm} | \Psi_{n}^{J^\pm} \rangle \Bigr\}\, C_n^{J^\pm} &= 0.
  \end{split}
   \label{eq:eigen}
\end{eqnarray}
}
\subsection{Variation of multiple AMD basis states}\label{sec:multi}

In the previous studies \cite{myo23b,myo25,tian24,cheng24,tian25},
we extended the cooling equation for the energy variation of the nuclear system with the multiple AMD basis states
and we call this method ``multicool method'' simply.

\red{We express the wave function $\Phi$ in the superposition of the intrinsic AMD basis states $\{\Phi_n\}$ with an index $n$ and a basis number $N_{\rm b}$}, and the total energy $E$ is given as
\begin{equation}
  \begin{split}
   \Phi&= \sum_{n=1}^{N_{\rm b}} C_n\,  \Phi_n , \quad
   E    = \dfrac{ \bra \Phi |H| \Phi \ket }{\bra \Phi | \Phi \ket}.
  \end{split}
  \label{eq:multi}
\end{equation}
In this method, the parity projection is performed and we omit the notation of parity $(\pm)$.

The multiple AMD configurations have variational parameters of $X_{n,i}=\{\bm{Z}_{n,i},\alpha^{\uparrow/\downarrow}_{n,i},C_n\}$.
The cooling equation to minimize the total energy $E$ is expressed as 
\begin{equation}
  \begin{split}
   \dfrac{{\rm d} X_{n,i}}{{\rm d} t}&= \frac{\mu}{\hbar} \dfrac{\partial E}{ \partial X_{n,i}^*},\quad \mbox{and c.c}.
  \end{split}
  \label{eq:multi_eq}
\end{equation}
By solving this equation, $\{X_{n,i}\}$ are obtained for the ground state.
It is noted that we do not impose any constraints on the spatial properties of the AMD configurations.

After this process, we generate the configurations of the excited states to be orthogonal to the ground-state configurations $\{\Phi_n\}$ in Eq.~(\ref{eq:multi}).
It is noted that if the configuration $\Phi_n$ is deformed to a specific direction,
the configuration orthogonal to $\Phi_n$ can be deformed to another direction, but
these configurations can have overlap after the angular momentum projection. 
To avoid this situation, we add the rotated configurations of $\{\Phi_n\}$ in the ground state.
We adopt two kinds of rotations: $(x,y,z)\to(z,x,y)$ and $(x,y,z)\to(y,z,x)$ according to the previous works \cite{myo23b,myo25}.
Finally, we define the ground-state configurations $\{\Phi_{c}\}$ with the index $c=1,\ldots,3N_{\rm b}$
and generate the excited-state configurations with the orthogonality to $\{\Phi_{c}\}$.

In this study, we adopt the orthogonal projection method \cite{isaka11,kukulin78,myo14}.
We define the pseudo potential $V_\lambda$ using $\{\Phi_{c}\}$ in the projection operator form with a strength $\lambda$.
We add $V_\lambda$ to the Hamiltonian and define $H_\lambda$ as
\begin{equation}
  \begin{split}
    H_\lambda &= H + V_\lambda,\quad
    V_\lambda = \lambda \sum_{c=1}^{3N_{\rm b}} \kets{\Phi_{c}}\bras{\Phi_{c}}.
  \end{split}
  \label{eq:PSE}
\end{equation}
\red{The wave function $\Phi_{\lambda}$ in the superposition of the basis states $\{\Phi_{\lambda,n}\}$ } and the total energy $E_{\lambda}$ are given as 
\begin{equation}
  \begin{split}
    \Phi_{\lambda} &= \sum_{n=1}^{N_{\rm b}} C_{\lambda,n} \Phi_{\lambda,n},\quad
    E_{\lambda} = \dfrac{ \bra \Phi_{\lambda} |H_{\lambda}| \Phi_{\lambda} \ket }{\bra \Phi_{\lambda} | \Phi_{\lambda} \ket}.
  \end{split}
\end{equation}
We perform the variation of $E_{\lambda}$ using Eq. (\ref{eq:multi_eq}) and determine $\Phi_{\lambda}$.
After the convergence of the solutions, we evaluate $E_{\lambda}$ omitting the contribution of the pseudo potential.

We begin to perform the energy variation with a small value of $\lambda$,
and repeat the variation gradually increasing $\lambda$.
When $\lambda$ becomes large, the wave function $\Phi_{\lambda}$ can be orthogonal to the configurations $\{\Phi_c\}$,
but each configuration $\Phi_{\lambda,n}$ is not imposed to be orthogonal to $\{\Phi_c\}$.
In a small value of $\lambda$, $\Phi_{\lambda}$ is not fully orthogonal to $\{\Phi_c\}$,
but the basis states $\{\Phi_{\lambda,n}\}$ can contribute to the low-lying nuclear states as well as the ground-state configurations.
Hence, we include them in the final basis states to be superposed.

We summarize the calculation step of the multicool method as follows.
\begin{enumerate}
\itemsep=2mm
\item[(1)]
  We prepare the initial AMD basis states randomly
  with the parameters $\{X_{n,i}\}$ for the basis index $n=1,\ldots,N_{\rm b}$ and the particle index $i=1,\ldots,A$.
  The initial coefficients $\{C_n\}$ are determined from the eigenvalue problem of the intrinsic Hamiltonian matrix.

\item[(2)]
  Solving the multicool equation in Eq. (\ref{eq:multi_eq}), we obtain the ground-state configurations $\{\Phi_{n}\}$.
  We further define $\{\Phi_{c}\}$ with $c=1,\ldots,3N_{\rm b}$ using the rotations of $\{\Phi_n\}$.
  
\item[(3)]
  Using the pseudo potential $V_\lambda$ in Eq.~(\ref{eq:PSE}),
  we solve the multicool equation and obtain the excited-state configurations $\{\Phi_{\lambda,n}\}$.

\item[(4)]
  We superpose $\{\Phi_n\}$ and $\{\Phi_{\lambda,n}\}$ with various $\lambda$ and
  solve the eigenvalue problem of the Hamiltonian matrix with the angular-momentum projection in Eq. (\ref{eq:eigen}). 
\end{enumerate}

In the calculation, the number of the basis states is typically around $N_{\rm b}=15-20$ in Eq.~(\ref{eq:multi}), 
and the total basis number in step 4 is at most about 150 for $^{20}$Ne \red{before the $K$ mixing.}

\red{
We discuss some properties of the multicool calculation with AMD.
In the results of the variation, some of the AMD configurations can have a similar intrinsic shape with different directions of rotation,
which can be seen in the previous studies \cite{myo23b,myo25}.
This is considered to be the role of the angular momentum projection, namely the restoration of the rotational symmetry.
In addition, we obtain the cluster configurations with the same constituents of clusters, showing the different relative distances between clusters.
This is considered to be the optimization of the relative wave function between clusters. }

\red{
To increase the speed of convergence in the variation including numerical time,
we can employ the initial AMD configurations, which are constructed with the constraints of the deformation and radius,
instead of the random configurations.
It is noted that the properties of the total wave function in step 4 are not changed. 
}

\red{
  It is also noted that the AMD wave function tends to give numerically the small determinant of the overlap matrix increasing the mass number $A$
  because of the antisymmetrization of nucleons having the Gaussian wave packets.
  In the present $^{20}$Ne case, we do not have this problem, but we carefully check the overlap matrix in the calculation.
  Recently, a new scheme was proposed to avoid this problem by introducing the superposition of Gaussians in the single-nucleon wave function
  \cite{kimura24}.
}
\subsection{Hamiltonian}\label{sec:ham}
In the Hamiltonian, we use the effective nuclear interactions,
which consist of the two-body central, spin-orbit, and Coulomb forces, and the repulsive three-body force with zero range for saturation of density.
In this study, we use MV1 (modified Volkov no.1) and G3RS for the spin-orbit force  \cite{ando80,kanada98,furutachi08}.
We use case 3 of MV1 with the same parameters as used in the previous works with the AMD model \cite{furutachi08};
Majorana parameter $M$=0.61 and 3000 MeV of the spin-orbit strength.
This setting reproduces well the systematic binding energies of $^4$He, $^{12}$C, and $^{16,18,20}$O \cite{furutachi08}.

Following the previous work \cite{furutachi08}, we set $\nu=0.17$ fm$^{-2}$ in Eq.~(\ref{eq:AMD}),
and the binding energy of $^4$He with the $(0s)^4$ configuration is 26.69 MeV.
In the multicool calculation, the binding energies of $^{12}$C and $^{16}$O are 91.17 and 129.08 MeV, respectively,
which agree with the experimental values of 92.16 and 127.62 MeV, respectively.
For $^{12}$C, the excitation energy of the $2^+_1$ state is obtained as 5.43 MeV, which is close to the experimental value of 4.44 MeV.

\section{Results}\label{sec:result}

\begin{table}[t]
\begin{center}
  \caption{
    Total energies and matter radii of the intrinsic states of $^{20}$Ne with positive and negative parities
    in the multicool calculation in comparison with the single AMD basis calculations.
  }
\label{tab:multi}
\renewcommand{\arraystretch}{1.5}
\begin{tabular}{lcccccc}
\noalign{\hrule height 0.5pt}
                              && \multicolumn{2}{c}{Positive parity } && \multicolumn{2}{c}{Negative parity}\\ \cline{3-4} \cline{6-7}
                              &&   Single      & Multicool            && Single       & Multicool \\ 
\noalign{\hrule height 0.5pt}
Energy (MeV)                  &&  $-149.7$     &  $-152.4$            &&  $-140.5$    &  $-143.3$ \\   
Radius (fm)                   &&   2.71        &      2.79            &&    2.99      &  2.88     \\
\noalign{\hrule height 0.5pt}
\end{tabular}
\end{center}
\end{table}

\begin{figure*}[th]
\centering
\includegraphics[width=7.5cm,clip]{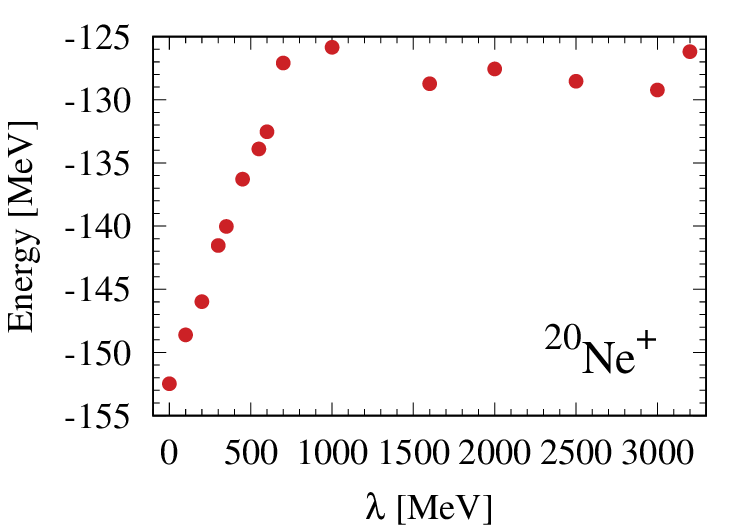}\hspace*{0.2cm}
\includegraphics[width=7.5cm,clip]{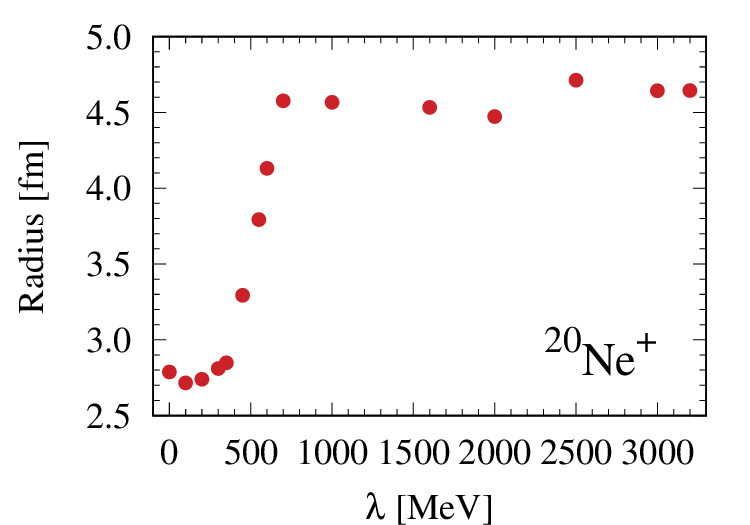}\hspace*{0.0cm}
\caption{
  Intrinsic energy (left) and radius (right) of the total wave function of $^{20}$Ne with positive parity state.
  The strength $\lambda$ of the pseudo potential changes.}
\label{fig:ene_positive}
\end{figure*}
\begin{figure*}[th]
\centering
\includegraphics[width=7.5cm,clip]{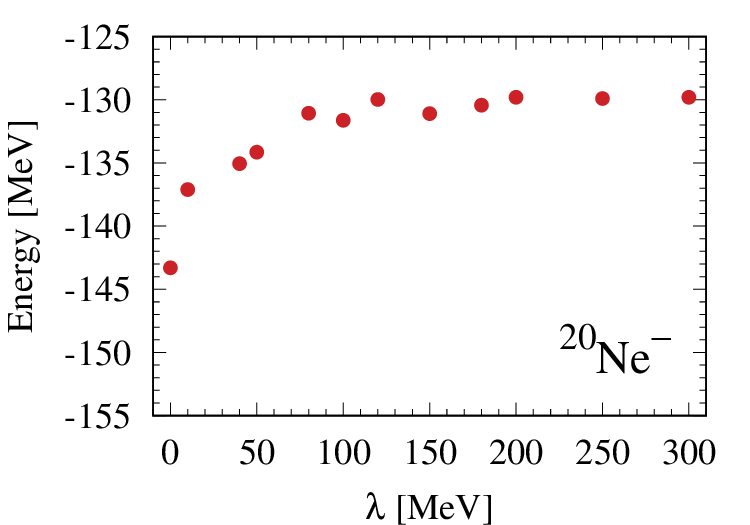}\hspace*{0.2cm}
\includegraphics[width=7.5cm,clip]{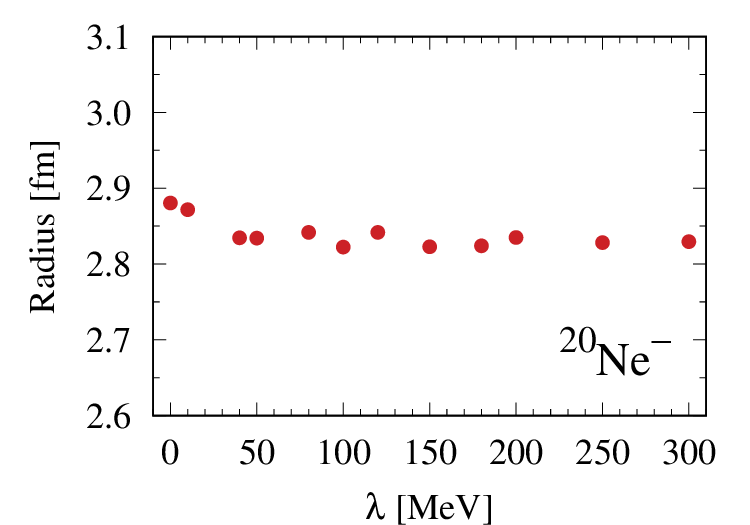}\hspace*{0.0cm}
\caption{
  Intrinsic energy (left) and radius (right) of the total wave function of $^{20}$Ne with negative parity state.
  The strength $\lambda$ of the pseudo potential changes.}
\label{fig:ene_negative}
\end{figure*}

\begin{figure}[bh]
\hspace*{-0.2cm}
\includegraphics[width=8.8cm,clip]{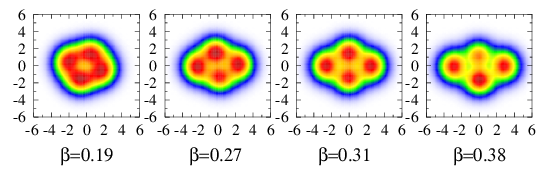}\\[2mm]
\hspace*{-0.2cm}
\includegraphics[width=8.8cm,clip]{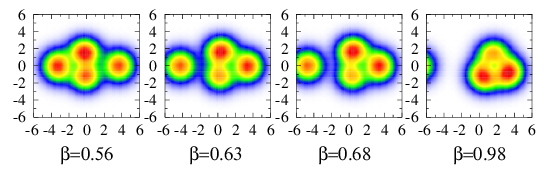}
\caption{
  \red{
  Intrinsic density distributions of the representative configurations of $^{20}$Ne for the positive parity state.
  The upper four panels (lower four panels) show the ground state (the excited state with $\lambda=1000$ MeV).
  Units of densities and axes are fm$^{-3}$ and fm, respectively.
  The deformation parameters $\beta$ of the configurations are shown at the bottom of each panel.
  }
  }
\label{fig:density0}
\end{figure}

\subsection{Energy variation} 

We perform the multicool calculation for $^{20}$Ne using the intrinsic AMD basis states with a number $N_{\rm b}=16$ in Eq.~(\ref{eq:multi}). 
The results of the energy variation in the intrinsic frame are summarized in Table \ref{tab:multi}.
For the positive parity state, the energy is $-152.4$ MeV and the radius is 2.79 fm.
For the negative parity state, the energy is $-143.3$ MeV and the radius is 2.88 fm.
We compare the results with those in a single AMD basis state.
The energy gain in the multiple bases is 2.7 MeV for positive parity and 2.8 MeV for the negative parity, which are similar to each other. 

In the matter radii, the negative parity state gives larger values than those of the positive parity state.
The effect of the multiple bases on the radii is about 0.1 fm in comparison with the single basis calculations in both parities,
but their roles are different.
An enhancement in the positive parity state indicates the addition of the deformed basis states including the $\alpha$ cluster states.
The reduction in the negative parity state indicates the addition of the spatially compact shell-like states.
Later, we discuss the spatial shapes and deformation of $^{20}$Ne in terms of the representative intrinsic density distributions.

Next, we generate the basis states for the excited states of $^{20}$Ne with the pseudo potential $V_\lambda$ keeping the basis number $N_{\rm b}=16$.
In Fig.~\ref{fig:ene_positive}, we show the results of the positive parity state of $^{20}$Ne
for total energy and matter radius by changing the strength $\lambda$ in the pseudo potential in Eq.~(\ref{eq:PSE}).
At $\lambda=0$, the calculation corresponds to the lowest-energy state in Table \ref{tab:multi}.
It is noted that the total energies are evaluated omitting the contribution of the pseudo potential.

By increasing $\lambda$, the energy gradually increases and becomes stable at around $-125$ to $-130$ MeV, which indicates the excited states.
The radius also increases to be around 4.5 fm, which is obviously expanded from the ground-state radius.
In the excited states, the $^{16}$O+$\alpha$ type configurations are often mixed, which is confirmed later from the intrinsic density distributions.

In Fig.~\ref{fig:ene_negative}, we show the results of the negative parity state of $^{20}$Ne.
By increasing $\lambda$, the energy gradually increases around $-130$ MeV.
It is found that the radius does not change so much at around 2.83 fm and is slightly reduced from the initial value of 2.88 fm,
This means that the nucleus is slightly shrunk, and we can confirm this effect from the density distributions in the excited states later.

  In Fig. \ref{fig:density0}, we show the intrinsic density distributions of the representative AMD configurations of $^{20}$Ne
  for the positive parity state, corresponding to Fig. \ref{fig:ene_positive}.
  The upper four panels are for the ground state and the lower four panels are for the excited state with $\lambda=1000$ MeV in Fig. \ref{fig:ene_positive}.
  We adjust the direction of the longest distribution to be the horizontal axis.
  We also show the quadrupole deformation parameter $\beta$ in each panel, the definition of which is given in Ref. \cite{dote97}.
  We obtain the various configurations with different $\beta$ deformations.
  In the excited-state configurations, the $\alpha$ clustering is enhanced with large $\beta$ by 
  changing the relative distance between $\alpha$ and $^{16}$O, where $^{16}$O has a tetrahedron like structure.

\clearpage

\subsection{Energy spectrum}

We superpose the AMD basis states obtained in Figs.~\ref{fig:ene_positive} and \ref{fig:ene_negative} in each parity with the angular momentum projection.
Solving the eigenvalue problem of the Hamiltonian matrix in Eq.~(\ref{eq:eigen}),
we obtain the ground state of $^{20}$Ne, the properties of which are summarized in Table \ref{tab:ground}.
The total energy and the radii of the ground state of $^{20}$Ne ($0^+_1$) are consistent with the experimental values.
The total energy is $-158.83$ MeV and the threshold energy of the $^{16}$O+$\alpha$ channel is $-155.77$ MeV.
Their difference is $3.06$ MeV, which is close to the experimental value of 4.73 MeV.

Experimentally, the charge radius is precisely observed \cite{angeli13}, and our result of 3.03 fm is close to, but slightly larger than the experimental value.
The calculated matter radius of 2.91 fm is also close to, but slightly larger than the range of the experimental analysis \cite{chulkov96}.
The AMD calculation with Gogny force \cite{isaka11} gives a slightly larger matter radius of 2.97 fm than the present value.
In the present calculation, the radii of protons and neutrons are 2.93 and 2.90 fm, respectively. 

If we use only the $N_{\rm b}=16$ basis states shown in Table \ref{tab:multi} for $^{20}$Ne,
the energy of the ground $0^+_1$ state is $-156.77$ MeV, indicating the energy gain from the basis states obtained with the pseudo potential is 2.06 MeV.

\begin{figure*}[th]
\centering
\includegraphics[width=14.5cm,clip]{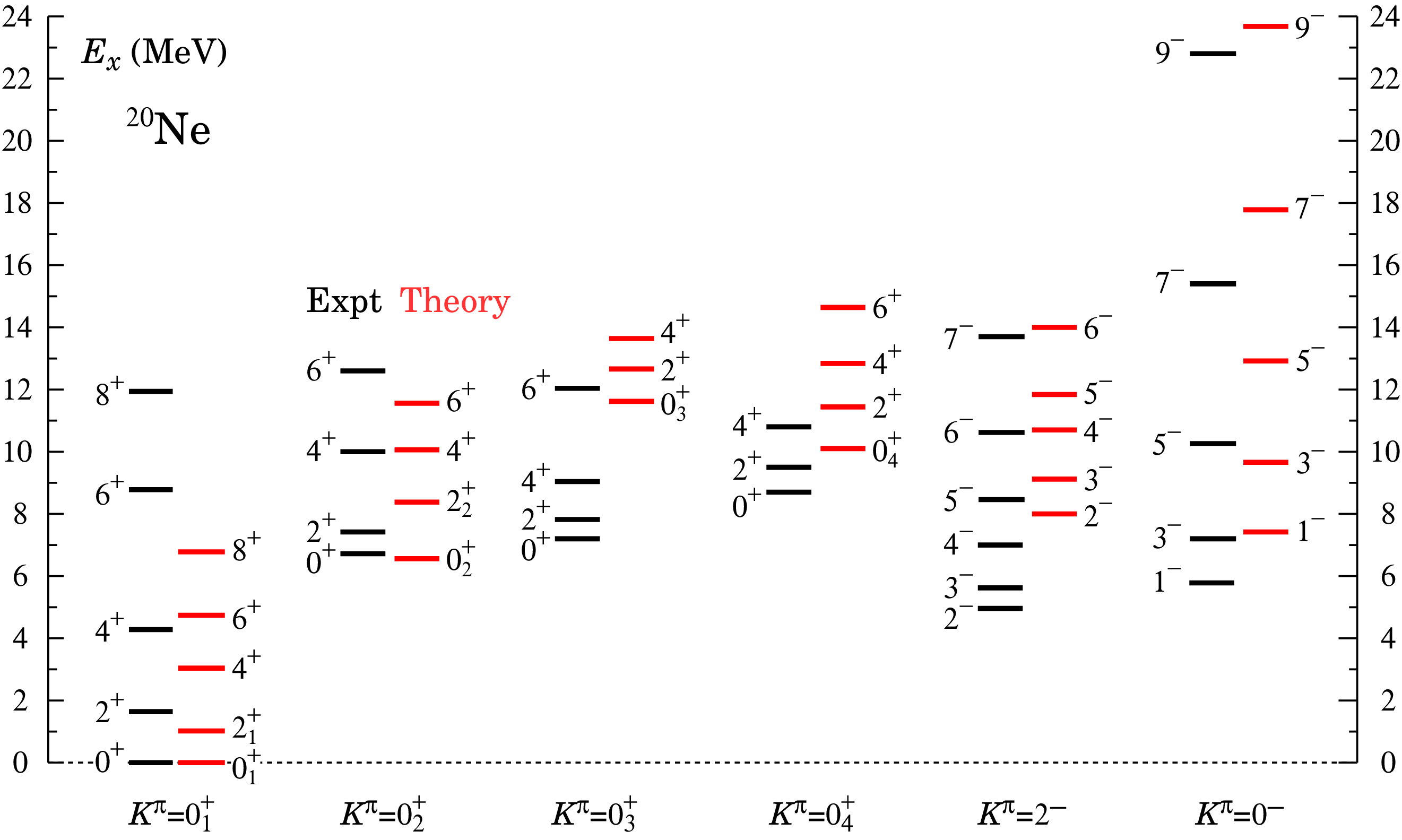}
\caption{
  Excitation energy spectrum of $^{20}$Ne for the experiment (black) \cite{tilley98} and the multicool calculation (red) in units of MeV.}
\label{fig:level}
\end{figure*}

\red{
We show the excitation energy spectrum of $^{20}$Ne in Fig. \ref{fig:level}, where the resonances are described within the bound-state approximation.
We identify the individual bands in terms of the behavior of the configuration mixing, radius, contribution of the spin-orbit force,
which is sensitive to the cluster configurations, and the transition strengths between the states.
Finally, we obtain the $K^\pi=0^+_{1-4}$ bands and the $K^\pi=0^-$ and $2^-$ bands.}
It is noted that the theoretical excitation energy of the $0^+_3$ state is higher than that of the $0^+_4$ state in Fig. \ref{fig:level},
but we assign this numbering considering the correspondence to the experimental levels and also the structural properties of $0^+_{3,4}$ states.
We discuss the properties of each band in detail later.

In Table \ref{tab:quant}, we list the properties of the bandhead states: the principal quantum number $N$ of the harmonic oscillator
and the matter radius, where $N$ is evaluated in terms of the expectation values of the kinetic energy and matter radius.
In $^{20}$Ne, the $0^+_1$ and $2^-$ states show the smallest group of $N$ with around 21--22, and
their values are close to the shell model limits of $N=20$ for positive-parity and $N=21$ for negative-parity states.
The resulting values are slightly above these limitations; hence, we can interpret that they are shell-like states. 
For the radius of the ground $0^+_1$ state, the configuration with $N=20$ gives $\sqrt{97/(80\nu)}$,
which leads to 2.67 fm, and the calculated radius of 2.91 fm is larger than this value by 0.24 fm.
This means the inclusion of the spatial correlations of the deformation or clustering beyond the $sd$ shell in the ground state.

For $0^+_2$, this state is known to be described in the $sd$-shell configurations \cite{tomoda78}, but is difficult in AMD \cite{kimura04}.
In the present multicool method, we obtain the corresponding state with the quanta $N=22.26$,
which suggests the dominant $sd$-shell component with a small mixing of the excitation components beyond the $sd$ shell.
The radius of 2.94 fm is close to that of the ground state.

For $0^+_3$, this state could be related to the $^{12}$C+$2\alpha$ configuration from the experimental analyses \cite{middleton71,nagatani71,greenwood75}.
In the present results, we obtain the state with the quanta $N=24.8$,
which is similar to that obtained in the AMD calculation with the $^{12}$C+2$\alpha$ constraint \cite{taniguchi04}.
The matter radius is 3.08 fm, which is larger than that of the ground state by 0.17 fm, and the same feature is reported in Ref. \cite{taniguchi04}.

For $0^+_4$, this state is considered to have the $^{16}$O+$\alpha$ structure as a higher-nodal state of the ground $0^+_1$ state.
We obtain the very large quanta of $N=40.38$ and also the large radius of 3.84 fm.
These values suggest the spatially developed cluster structure of $^{20}$Ne.
Later we discuss the density distribution of the representative configurations on this state.

For the $1^-$ state, the quanta value is $N=26.62$ and the radius is 3.24 fm, which are larger values than those of the ground $0^+_1$ state.
This state also shows the spatially developed structure of the $^{16}$O+$\alpha$ configuration, but not as large as that of the $0^+_4$ state.

For the $2^-$ state, the quanta value is $N=21.99$ and the radius is 2.89 fm, which are smaller values than those of the $1^-$ state
and are also close to the shell-model states of $0^+_{1,2}$.
The radial properties of the $1^-$ and $2^-$ states are the same as those reported in the AMD calculations \cite{isaka11}.
Next, we discuss the properties of each of the $K^\pi$ band states in detail.

\begin{table}[t]
\begin{center}
  \caption{
    Ground-state properties of $^{20}$Ne ($0^+_1$); the total energy in units of MeV and
    the radii of matter ($r_{\rm m}$), charge ($r_{\rm ch}$), proton ($r_{\rm p}$), and neutron ($r_{\rm n}$) in units of fm,
    in comparison with the experimental analyses \cite{angeli13,chulkov96}.
  }
\label{tab:ground}
\renewcommand{\arraystretch}{1.5}
\begin{tabular}{lcllllllll}
\noalign{\hrule height 0.5pt}
         &  Energy          && $r_{\rm m}$      && $r_{\rm ch}$ & $r_{\rm p}$ & $r_{\rm n}$ \\
\noalign{\hrule height 0.5pt}
Present  &  $-158.83$       && 2.91             && 3.03         & 2.93        & 2.90 \\
Expt     &  $-160.64$       && 2.84(3)--2.87(3) && 3.0055(21)   &             &      \\
\noalign{\hrule height 0.5pt}
\end{tabular}
\end{center}
\end{table}

\begin{table}[t]
\begin{center}
  \caption{
    Principal quantum numbers $N$ of $^{20}$Ne for the six bandhead states with the assignment of $K^\pi$
    and their matter radii in units of fm in the multicool calculation, in comparison with those in AMD \cite{isaka11}.
  }
\label{tab:quant}
\renewcommand{\arraystretch}{1.5}
\begin{tabular}{lcccccccccccc}
\noalign{\hrule height 0.5pt}
$J^\pi$            && $0^+_1$ && $0^+_2$ && $0^+_3$ && $0^+_4$ && $1^-$  && $2^-$ \\
$K^\pi$            && $0^+_1$ && $0^+_2$ && $0^+_3$ && $0^+_4$ && $0^-$  && $2^-$ \\
\noalign{\hrule height 0.5pt}
$N$                && 21.38   && 22.26   &&  24.81  && 40.38   && 26.62  && 21.99 \\
Radius             && 2.91    &&  2.94   &&   3.08  &&  3.84   &&  3.24  && 2.89  \\
AMD \cite{isaka11} && 2.97    &&         &&         &&         &&  3.27  && 2.98  \\
\noalign{\hrule height 0.5pt}
\end{tabular}
\end{center}
\end{table}

\subsection{$K^\pi=0^+_1$}

We discuss the $K^\pi=0^+_1$ band states.
We obtain the five states with $0^+$, $2^+$, $4^+$, $6^+$, and $8^+$ as shown in Fig. \ref{fig:level}.
It is found that the level spacing is smaller than the experiments, indicating the large value of the moment of inertia,
which is the same feature obtained in the previous AMD calculations \cite{kimura04,isaka11,taniguchi04}.

\begin{figure}[b]
\hspace*{-0.2cm}
\includegraphics[width=8.8cm,clip]{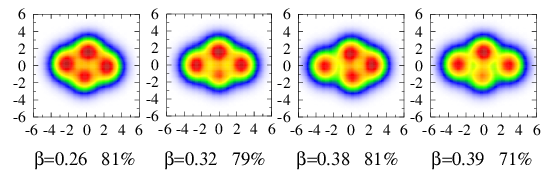}
\caption{
  Intrinsic density distributions of the representative configurations of $^{20}$Ne ($0^+_1$) in the $K^\pi=0^+_1$ band.
  Units of densities and axes are fm$^{-3}$ and fm, respectively.
  \red{The deformation parameters $\beta$ and the squared overlaps with the $0^+_1$ state in units of \% are also shown at the bottom of each panel.}
}
\label{fig:density1}
\end{figure}

The intrinsic density distributions of the representative AMD configurations in the $0^+_1$ state are shown in Fig. \ref{fig:density1}
with the quadrupole deformation parameter $\beta$.
\red{We also show the squared overlaps of the configurations with respect to the total $0^+_1$ wave function.}
It is found that the deformation is induced in association with the $\alpha$ clustering.
Some densities show the $^{16}$O+$\alpha$ configuration with a tetrahedron-like structure of $^{16}$O,
and the distance between $^{16}$O and $\alpha$ depends on the configurations.
The deformation parameter $\beta$ is dominantly around 0.4, and this result agrees with the analysis with AMD using the deformed Gaussian wave packet,
in which the energy minimum of the $0^+_1$ configuration is obtained at around $\beta=0.4$ \cite{kimura04}.

The electric quadrupole transitions $B(E2)$ in this band are shown in Table \ref{tab:E2_1st} in comparison with the experiments and other theories;
the $(sd)^4$ shell model (SM) with the effective charge of $\delta e=0.54e$ \cite{tomoda78},
resonating group method (RGM) of $^{16}$O+$\alpha$ \cite{matsuse75},
coupled-channel orthogonality condition model (OCM) of $^{16}$O+$\alpha$ and $^{12}$C+$^8$Be \cite{fujiwara78},
generator coordinate method (GCM) of (4$\alpha$)+$\alpha$ \cite{dufour94},
AMD+GCM \cite{kimura04},
\red{no-core symplectic shell model (NCSpM) \cite{tobin14},}
and the relativistic Hartree-Bogoliubov method (RHB) \cite{marevic18}.

It is found that the global tendency is reproduced and is similar to other theories.
Our results give rather smaller values than other theories, in particular, for higher spin cases,
which suggests more quadrupole deformation in the present wave function.
One of the possible extensions is the adoption of the deformed Gaussian wave packets of the nucleon wave function in Eq.~(\ref{eq:AMD})
in the present multicool calculation, as introduced in Refs. \cite{kimura04,isaka11,taniguchi04}. 

The electric quadrupole moment $Q$ of the $2^+_1$ state is shown in Table \ref{tab:Q}.
Our result is similar to those of other theories and belongs to the group of the smaller value.
The experimental value is the adopted one, and it is known that this value is larger than theoretical ones \cite{stone05,spear81},
which is a long-standing problem.

\begin{table*}[t]
\begin{center}
  \caption{
    Electric quadrupole transitions $B(E2)$ of $^{20}$Ne in the $K^\pi=0^+_1$ band in units of $e^2$fm$^4$
    in comparison with the experiments \cite{tilley98,tunl,nndc} and other theories.
    The effective charge is $\delta e=0.54e$ in the shell model (SM).
  }
\label{tab:E2_1st}
\renewcommand{\arraystretch}{1.5}
\begin{tabular}{lccccccccc}
\noalign{\hrule height 0.5pt}  \vspace*{-0.15cm}
                  &  Expt     & Present & SM              & RGM              & OCM               &  GCM            & AMD             & \red{NCSpM}    & RHB              \\
                  &           &         & \cite{tomoda78} & \cite{matsuse75} & \cite{fujiwara78} & \cite{dufour94} & \cite{kimura04} & \cite{tobin14} & \cite{marevic18} \\
\noalign{\hrule height 0.5pt}
$2^+_1 \to 0^+_1$ & 65$\pm$3  & 41.2    &  57.0           &  36.2            &  33.2             &  50.0            & 70.3           & 59.7    & 54  \\
$4^+   \to 2^+_1$ & 71$\pm$6  & 42.0    &  69.9           &  45.2            &  41.3             &  64.4            & 83.7           & 79.3    & 89  \\
$6^+   \to 4^+  $ & 64$\pm$10 & 25.6    &  57.9           &  36.5            &  33.3             &  55.3            & 52.7           & 74.2    & 85  \\
$8^+   \to 6^+  $ & 29$\pm$4  & 12.0    &  35.5           &  19.7            &  20.3             &  ---             & 21.0           & --      & --  \\
\noalign{\hrule height 0.5pt}
\end{tabular}
\end{center}
\end{table*}
\begin{table*}[t]
\begin{center}
  \caption{
    Electric quadrupole moment $Q$ of $^{20}$Ne ($2^+_1$) in units of $e$\,fm$^2$.
    The effective charge is $\delta e=0.54e$ in the shell model (SM).
  }
\label{tab:Q}
\renewcommand{\arraystretch}{1.5}
\begin{tabular}{lcccccccc}
\noalign{\hrule height 0.5pt}  \vspace*{-0.15cm}
Expt                   & Present  & SM              & RGM               & OCM               & OCM            & GCM            & \red{NCSpM}   & RHB \\
\cite{stone05,spear81} &          & \cite{tomoda78} & \cite{matsuse75}  & \cite{fujiwara78} & \cite{kazama86}& \cite{dufour94}& \cite{tobin14}& \cite{marevic18}  \\
\noalign{\hrule height 0.5pt}                                                                               
$-23\pm 3$             & $-12.11$ &  $-15.5$        &  $-12.14$         & $-11.6$           & $-15.95$       & $-14.3$        & $-15.69$      & $-16.61$ \\ 
\noalign{\hrule height 0.5pt}
\end{tabular}
\end{center}
\end{table*}

\subsection{$K^\pi=0^+_2$}

We discuss the $K^\pi=0^+_2$ band states.
The excitation energy of the $0^+_2$ state is 6.55 MeV in the multicool calculation,
which agrees with the experimental value of 6.725 MeV \cite{tilley98}, as shown in Fig.~\ref{fig:level}.
If we use only the $N_{\rm b}=16$ basis states without the pseudo potential,
the excitation energy of $0^+_2$ is 7.38 MeV, and
the superposition of the additional basis state with the pseudo potential gives the gain of the excitation energy by 0.83 MeV.

\begin{figure}[b]
\hspace*{-0.2cm}
\includegraphics[width=8.8cm,clip]{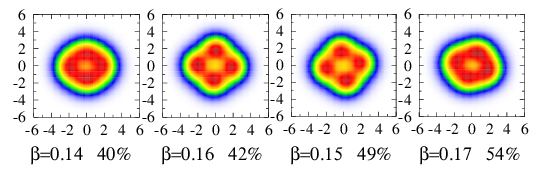}
\caption{
  Intrinsic density distributions of the representative configurations of $^{20}$Ne ($0^+_2$) in the $K^\pi=0^+_2$ band.
  Units of densities and axes are fm$^{-3}$ and fm, respectively.
  \red{
  The deformation parameters $\beta$ and the squared overlaps with the $0^+_2$ state are also shown at the bottom of each panel.}
}
\label{fig:density2}
\end{figure}

The intrinsic density distributions of the representative AMD configurations in the $0^+_2$ state are shown in Fig. \ref{fig:density2}.
These densities commonly indicate less deformation, and the $\beta$ parameters show the small value of around $\beta=0.15$.
A small portion of the $\alpha$ clustering is confirmed, which causes the deviation from the spherical shape.
In the second and third panels, four $\alpha$ clusters can be confirmed with close distances,
and the last one is located at the origin and perpendicular to the plane where the four $\alpha$ particles lie, like the pyramid shape.

\red{
It is an interesting fact that the $0^+_2$ state is not obtained in the previous AMD calculations with the $\beta$ constraint \cite{kimura04,taniguchi04,isaka11}, in which the basis states are generated at the energy-minimum point with respect to the specific $\beta$ deformation.
However, it is not obvious that these basis states are suitable in the description of the $0^+_2$ state.
This fact suggests that the $0^+_2$ state may not be an energy-minimum state with respect to the specific $\beta$ deformation,
which could be the reason for the difficulty to generate the appropriate basis states for the $0^+_2$ state.}

In the present multicool calculation, we generate the multiple basis states without assuming any deformation,
This framework reasonably works to generate the basis states necessary for the $0^+_2$ state.
In addition, the pseudo potential with a small strength of $\lambda$ contributes to converge the results
because we can optimize the basis states for the excited states without the constraints on the specific shape of the nucleus.

The electric quadrupole transitions associated with the $0^+_2$ and $2^+_2$ states in $K^\pi=0^+_2$ are shown in Table \ref{tab:E2_2nd}.
The strength of $0^+_2 \to 2^+_1$ in the present calculation is larger than that of the experimental value, but give the same order.
Other theories give small values in comparison with the experiment.

The strength of $2^+_2 \to 0^+_1$ in the present calculation is similar to the OCM calculation \cite{fujiwara78} and larger than the experimental value,
although the magnitude is small commonly in theory and experiment.
The strength of $2^+_2 \to 2^+_1$ in the present calculation is consistent with the experimental value.

The magnitudes of the matrix elements of the electric monopole transition $|M(E0)|$ from the ground $0^+_1$ state are shown in Table \ref{tab:E0}.
Our results fairly reproduce the experimental values for the $0^+_{2,3}$ states \cite{mitsunobu72}.
We also show the values of the isoscalar monopole transition (IS0), which almost give twice those of the electric cases.

%
\begin{table}[t]
\begin{center}
  \caption{
    Electric quadrupole transitions $B(E2)$ of $^{20}$Ne associated with $0^+_2$ and $2^+_2$ in $K^\pi=0^+_2$
    and also the $2^+_1$ state in $K^\pi=0^+_1$ in units of $e^2$fm$^4$ in comparison with the experiments \cite{tilley98,tunl,nndc,fujiwara78}.
    The effective charge is $\delta e=0.54e$ in the shell model (SM).
  }
\label{tab:E2_2nd}
\renewcommand{\arraystretch}{1.5}
\begin{tabular}{lccccccc}
\noalign{\hrule height 0.5pt} \vspace*{-0.15cm}
                  &  Expt           & Present & SM                  & OCM              \\
                  &                 &         & \cite{tomoda78}     & \cite{fujiwara78}\\
\noalign{\hrule height 0.5pt}
$0^+_2 \to 2^+_1$ & 13.5$\pm$3.6    & 27.7    &  2.5                & 0.29 \\
$0^+_3 \to 2^+_1$ &  1.0$\pm$0.2    & 0.75    &                     & 3.52 \\ 
$0^+_4 \to 2^+_1$ &                 & 3.27    &                     &      \\
$2^+_2 \to 0^+_1$ &  0.17$\pm$0.03  & 0.60    & $9.1\times 10^{-4}$ & 0.74 \\
$2^+_2 \to 2^+_1$ &  5.45$\pm$0.69  & 7.33    & 1.2                 & 1.41 \\
$2^+_2 \to 4^+_1$ &  $<8.6$         & 0.35    & 0.21                & 0.53 \\
\noalign{\hrule height 0.5pt}
\end{tabular}
\end{center}
\end{table}

%
\begin{table}[t]
\begin{center}
  \caption{
    Magnitudes of the matrix elements of the electric monopole transitions $|M(E0)|$ of $^{20}$Ne
    from the ground $0^+_1$ state in units of $e$\,fm$^2$ in comparison with the experiments \cite{mitsunobu72}.
    The isoscalar monopole transitions $|M({\rm IS}0)|$ are also shown in the calculation in units of fm$^2$.
  }
\label{tab:E0}
\renewcommand{\arraystretch}{1.5}
\begin{tabular}{lcccccc}
\noalign{\hrule height 0.5pt}
                  &  \multicolumn{2}{c}{$E0$} && IS0  \\ \cline{2-3} \cline{5-5}
                  &  Expt         & Present   && Present \\ 
\noalign{\hrule height 0.5pt}
$0^+_1 \to 0^+_2$ & 7.37$\pm$1.97 & 6.26      && 12.51   \\ 
$0^+_1 \to 0^+_3$ & 6.90$\pm$1.44 & 4.47      && ~8.55   \\ 
$0^+_1 \to 0^+_4$ &               & 4.68      && ~9.46   \\ 
\noalign{\hrule height 0.5pt}
\end{tabular}
\end{center}
\end{table}

\subsection{$K^\pi=0^+_3$}

We discuss the $K^\pi=0^+_3$ band states.
This band is suggested to be related to the $^{12}$C+$2\alpha$ configuration \cite{middleton71,nagatani71,greenwood75},
although the experimental binding energy of the bandhead $0^+_3$ state is 4.7 MeV measured from the $^{12}$C+$2\alpha$ threshold,
which is not small according to the threshold rule \cite{ikeda68}.

In Fig. \ref{fig:level}, we assign the state at the excitation energy of 11.61 MeV to the $0^+_3$ state,
considering the quanta $N$, the radius as shown in Table \ref{tab:quant}, and the representative configurations,
according to the previous study with AMD introducing the $^{12}$C+$2\alpha$ configurations \cite{taniguchi04}.

\begin{figure}[t]
\hspace*{-0.2cm}
\includegraphics[width=8.8cm,clip]{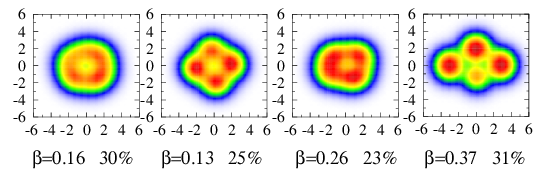}
\caption{
  Intrinsic density distributions of the representative configurations of $^{20}$Ne ($0^+_3$) in the $K^\pi=0^+_3$ band.
  Units of densities and axes are fm$^{-3}$ and fm, respectively.
  \red{The deformation parameters $\beta$ and the squared overlaps with the $0^+_3$ state are also shown at the bottom of each panel.}
}
\label{fig:density3}
\end{figure}

In Fig. \ref{fig:density3}, we show the intrinsic density distributions of the representative configurations of the $0^+_3$ state.
It is found that the third panel corresponds to the $^{12}$C+$2\alpha$ configuration,
although it does not show a clear cluster structure of $^{12}$C+$2\alpha$.
In the configuration, $2\alpha$s are located at the right-hand side and $^{12}$C is located at the left-hand side
with the oblate shape, where its deformation plane is almost perpendicular to the presenting plane.
The fourth panel indicates the $\alpha$-$^{12}$C-$\alpha$ configuration, where a middle $^{12}$C shows a $3\alpha$-like structure and
its deformation plane is also perpendicular to the presenting plane.

The obtained excitation energy of 11.61 MeV for the $0^+_3$ state is higher than the experimental value of 7.19 MeV by 4.42 MeV.
We consider the reason for it as follows:
In the present calculation, the energy of $^{12}$C ($0^+_1$) is $-91.17$ MeV after the superposition with angular momentum projection.
However, when we generate the basis states of $^{20}$Ne in the multicool method, 
the intrinsic energy of the single configuration of $^{12}$C is $-81.98$ MeV and underestimated from that of the projected $^{12}$C by about 9 MeV.
In addition, the superposition of the $^{12}$C configurations in the $^{20}$Ne basis states is difficult to treat in the present method.
Hence, we expect that the sufficient energy of $^{12}$C is not obtained in the present $^{20}$Ne calculation.
This is the possible reason why the calculated excitation energy of the $0^+_3$ state is higher than the experimental one.
\red{In fact, the squared overlaps of the configurations in Fig. \ref{fig:density3} are not large and at most around 30\%,
which indicates that the single AMD configuration is not sufficient and the superposition of the configurations is necessary in the description of the $0^+_3$ state.}
The same problem is pointed out in the $\alpha$ cluster model \cite{itagaki16}.

\red{In Ref. \cite{taniguchi04}, the authors impose the constraint to generate $^{12}$C+$2\alpha$ configurations in the energy variation}
and also adopt the deformed Gaussian wave packet.
These effects nicely work to describe the $0^+_3$ state with a proper excitation energy.
It would be interesting to apply the deformed Gaussian wave packet to the present multicool method.

\subsection{$K^\pi=0^+_4$}

The $K^\pi=0^+_4$ band has the cluster structure of $^{16}$O+$\alpha$ and corresponds to a higher-nodal state of the $K^\pi=0^+_1$ band.
The excitation energy of the $0^+_4$ state is 10.11 MeV, which is slightly higher than the experimental value of 8.70 MeV,
but similar to that of the AMD calculation \cite{taniguchi04}.

\begin{figure}[t]
\hspace*{-0.2cm}
\includegraphics[width=8.8cm,clip]{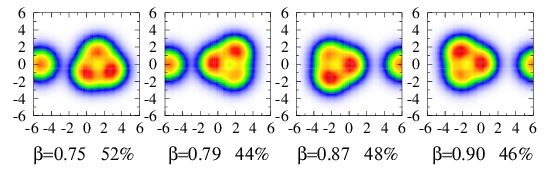}
\caption{
  Intrinsic density distributions of the representative configurations of $^{20}$Ne ($0^+_4$) in the $K^\pi=0^+_4$ band.
  Units of densities and axes are fm$^{-3}$ and fm, respectively.
  \red{The deformation parameters $\beta$ and the squared overlaps with the $0^+_4$ state are also shown at the bottom of each panel.}
}
\label{fig:density4}
\end{figure}

In Fig. \ref{fig:density4}, we show the intrinsic density distributions of the representative configurations of the $0^+_4$ state,
which clearly show the $^{16}$O+$\alpha$ cluster structure with a large relative distance between clusters and also a large $\beta$ value of around 0.8,
leading to the large matter radius of 3.84 fm as shown in Table \ref{tab:quant}.
\red{The relative distance between $^{16}$O and $\alpha$ depends on the configurations
  and this aspect contributes to optimize the relative wave function between clusters.
}

\red{
  It is noted that the $0^+_4$ band is obtained in a rather high excitation energy as shown in Fig. \ref{fig:level},
  and this comes from the orthogonality to the ground-state configurations.
  The same result is obtained in the $^{16}$O+$\alpha$ OCM calculation \cite{kruppa90},
  in which the resonances are described correctly using complex scaling.
  In Ref. \cite{kruppa90}, the authors show the Coulomb barrier as a function of the relative distance between clusters,
  which can stabilize the resonances. The same effect would occur in the present calculation.
}
%
\subsection{$K^\pi=0^-$}

The $K^\pi=0^-$ band is considered to be the parity doublet of the $K^\pi=0^+$ band \cite{fujiwara80}.
In the present calculation, the excitation energy of the $1^-$ state is 7.42 MeV, which is similar to the other theory of 7.0 MeV \cite{kimura04},
and is slightly higher than the experimental value of 5.79 MeV.

\begin{figure}[th]
\hspace*{-0.2cm}
\includegraphics[width=8.8cm,clip]{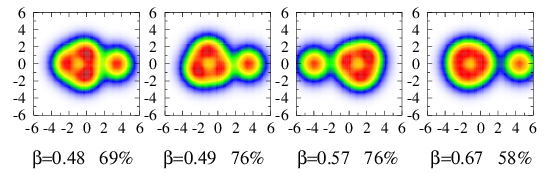}
\caption{
  Intrinsic density distributions of the representative configurations of $^{20}$Ne ($1^-$) in the $K^\pi=0^-$ band.
  Units of densities and axes are fm$^{-3}$ and fm, respectively.
  \red{
  The deformation parameters $\beta$ and the squared overlaps with the $1^-$ state are also shown at the bottom of each panel.}
}
\label{fig:density5}
\end{figure}

In Fig. \ref{fig:density5}, we show the intrinsic density distributions of the representative configurations of the $1^-$ state,
which dominantly have the $^{16}$O+$\alpha$ cluster structure with relatively large $\beta$ value of around 0.5--0.7,
but smaller than that of the $0^+_4$ state.
The matter radius of the $1^-$ state is 3.16 fm, as shown in Table \ref{tab:quant}, which is larger than that of the ground $0^+_1$ state,
but smaller than that of the $0^+_4$ state.

The electric quadrupole transitions in the $K^\pi=0^-$ band states are shown in Table \ref{tab:E2_3rd},
which are relatively larger values than those of the $K^\pi=0^+_1$ band in Table \ref{tab:E2_1st} due to the large deformation and radius.
The strengths in the present model are similar to other calculations and agree with the experimental value of the $3^- \to 1^-$ transition.

\begin{table}[t]
\begin{center}
  \caption{
    Electric quadrupole transitions $B(E2)$ of $^{20}$Ne in the $K^\pi=0^-$ band in units of $e^2$fm$^4$.
    in comparison with the experiments and other theories.
  }
\label{tab:E2_3rd} 
\renewcommand{\arraystretch}{1.5}
\begin{tabular}{lccccccc}
\noalign{\hrule height 0.5pt}    \vspace*{-0.15cm}
              &  Expt       & Present & RGM               & GCM             & AMD    & RHB \\
              &             &         & \cite{matsuse75}  & \cite{dufour94} & \cite{kimura04} & \cite{marevic18} \\
\noalign{\hrule height 0.5pt}
$3^- \to 1^-$ & 164$\pm$26  & 151.7   & 121.4 & 155  & 151.2  & 156 \\
$5^- \to 3^-$ &             & 187.6   & 133.4 & 206  & 182.4  & 177 \\
$7^- \to 5^-$ &             & 153.7   & 122.7 &      & 141.6  & 181 \\
$9^- \to 7^-$ &             & 107.7   &  67.8 &      &  87.9  &     \\
\noalign{\hrule height 0.5pt}
\end{tabular}
\end{center}
\end{table}

\subsection{$K^\pi=2^-$}

The $K^\pi=2^-$ band is considered to be the shell-like state
according to the properties of the bandhead $2^-$ state, as shown in Table \ref{tab:quant}, similar to the ground $0^+_1$ state.
One difference from the ground $0^+_1$ state is the attraction from the spin-orbit force is very large as $-21.83$ MeV in the $2^-$ state,
which is almost twice that of $-11.50$ MeV of the $0^+_1$ state.
From this property and the quanta of $N=21.99$,
we expect that in this $2^-$ state, one nucleon in the $0p_{1/2}$ orbit can be excited to the $0d_{5/2}$ orbit
to increase the contribution of the spin-orbit force, as the 5$p$-1$h$ configuration upon the $^{16}$O nucleus.

The excitation energy of the $2^-$ state is 8.00 MeV, which is higher than the experimental value of 4.97 MeV by 3 MeV.
In order to improve this energy difference,
the effect of the deformed mean field is discussed using the deformed Gaussian wave packet \cite{kimura04},
in which the excitation energy is about 7.0 MeV and slightly lower than our results, but still higher than the experiment.

\begin{figure}[t]
\hspace*{-0.2cm}
\includegraphics[width=8.8cm,clip]{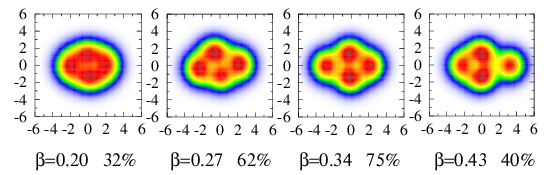}
\caption{
  Intrinsic density distributions of the representative configurations of $^{20}$Ne ($2^-$) in the $K^\pi=2^-$ band.
  Units of densities and axes are fm$^{-3}$ and fm, respectively.
  \red{
  The deformation parameters $\beta$ and the squared overlaps with the $2^-$ state are also shown at the bottom of each panel.}
}
\label{fig:density6}
\end{figure}

In Fig. \ref{fig:density6}, we show the intrinsic density distributions of the representative configurations of the $2^-$ state,
which show a rather small deformation with $\beta\simeq 0.3$.
In Ref. \cite{kimura04}, the energy minimum is obtained at around $\beta\simeq 0.4$, which is slightly larger than that of the present calculation.

The electric quadrupole transitions in the $K^\pi=2^-$ band are shown in Table \ref{tab:E2_4th},
which are relatively smaller values than the experiments and other theories, and the relative strengths of the transitions are fairly reproduced.
It is interesting to include the effect of the deformed mean-field in the present calculation.

\begin{table}[t]
\begin{center}
  \caption{
    Electric quadrupole transitions $B(E2)$ of $^{20}$Ne in the $K^\pi=2^-$ band in units of $e^2$fm$^4$.
    in comparison with the experiments and other theories.
  }
\label{tab:E2_4th} 
\renewcommand{\arraystretch}{1.5}
\begin{tabular}{lccccccc}
\noalign{\hrule height 0.5pt}     \vspace*{-0.15cm}
              &  Expt.            & Present & OCM   & AMD   \\
              &                   &         & \cite{fujiwara78} & \cite{kimura04} \\
\noalign{\hrule height 0.5pt}
$3^- \to 2^-$ & 111$\pm$29        & 47.9    & 83.1  & 102.8 \\
$4^- \to 3^-$ &  77$\pm$16        & 31.6    & 59.6  &  77.8 \\
$4^- \to 2^-$ &  34$\pm$6         & 19.8    & 26.3  & 38.5  \\
$5^- \to 4^-$ &  $<$808           & 24.0    & 34.9  & 84.5  \\
$5^- \to 3^-$ &  84$\pm$19        & 29.1    & 38.2  & 56.6  \\
$6^- \to 5^-$ &  32$\pm$13        & 14.9    & 26.2  & 29.9  \\
$6^- \to 4^-$ &  55$^{+23}_{-13}$ & 30.7    & 38.9  & 64.0  \\
\noalign{\hrule height 0.5pt}
\end{tabular}
\end{center}
\end{table}

\section{Summary}\label{sec:summary}

We proposed a new framework to construct the optimal configurations for nuclear many-body system
in the multiple Slater determinants of the antisymmetrized molecular dynamics (AMD).
The AMD configurations are superposed and determined simultaneously to minimize the energy of the total nuclear system.
We further optimize the excited-state configurations imposing the orthogonal condition to the ground-state configurations.
In this paper, we applied this method to $^{20}$Ne and focus on the structures of the shell-like and cluster states.

In AMD, the nucleon wave function is a Gaussian wave packet and
the centroid parameters of Gaussians play the decisive role in the descriptions of the shell and cluster structures of nuclei.
These parameters are determined in the cooling method for the multiple AMD basis states without the constraints on the shapes of the nucleus.
We call this method multicool or multiple cooling.

For $^{20}$Ne, there are many studies to investigate the low-lying structures of this nucleus.
So far, no microscopic theory can describe the six bands of $K^\pi=0^+_{1-4}, 0^-$, and $2^-$ comprehensively.
This difficulty is related to the variety of the structures of $^{20}$Ne, such as the shell-like, deformation, and clustering states.
The AMD method is a promising one to describe these band states; however, the $K^\pi=0^+_2$ band,
which is considered to be the shell-like states, cannot be obtained in the quadrupole-deformation constraints. 

In the multicool calculation, we consistently obtain six kinds of bands of $^{20}$Ne.
In particular, the $K^\pi=0^+_2$ band is nicely described as a shell-like state.
This is because, in the multicool method, the multiple AMD basis states are optimized without any physical constraints such as deformation.
This property is also beneficial to optimize the configurations for the excited states.
As a result, the $K^\pi=0^+_2$ band states are obtained with rather spherical shapes in the proper excitation energies,
and we can discuss the electric transitions associated with the $0^+_2$ and $2^+_2$ states.
The present results also mean that this band might be difficult to obtain with the constraints on the conventional quadrupole deformation.

In the multicool calculation, the excitation energies of some of the bandhead states are still higher than the experimental values,
such as in $K^\pi=0^+_3$ and $2^-$. 
This property might be related to the underestimation of the deformed mean-field effect in $^{20}$Ne, including the $^{12}$C description in the $^{20}$Ne configurations,
because the spherical Gaussian wave packet is employed for nucleons in the present AMD wave function. 
It would be interesting to adopt the deformed Gaussians \cite{kimura04}
to treat the deformed mean field effectively in the present multicool method.

\section*{Acknowledgments}
This work was supported by Japan Society for the Promotion of Science KAKENHI Grants No. JP21K03544, No. JP22K03643, and No. JP25H01268
and JapanScience and Technology Agency ERATO Grant No. JPMJER2304.
This work was also partly supported by the Research Center for Nuclear Physics Collaboration Research Network program
at Osaka University under Project No. COREnet-059.
M.L. acknowledges support from the National Natural Science Foundation of China (Grant No. 12105141)
and the Jiangsu Provincial Natural Science Foundation (Grant No. BK20210277).
Numerical calculations were partly achieved through the use of SQUID at D3 Center, Osaka University.

\bibliographystyle{apsrev4-2} 
\bibliography{reference_ne}  

\end{document}